\documentclass[pra,twocolumn,showpacs]{revtex4-1}

\usepackage{amsmath}
\usepackage{amsfonts}
\usepackage{amssymb}
\usepackage{mathrsfs}
\usepackage{latexsym}
\usepackage{dsfont}
\usepackage{graphicx}

\begin{document}

\title{Entanglement dynamics via geometric phases in quantum spin chains} 

\author{C. S. \surname{Castro}}
\email{ccastro@if.uff.br}
\author{M. S. \surname{Sarandy}}
\email{msarandy@if.uff.br}

\affiliation{Instituto de F\'{\i}sica, Universidade Federal Fluminense, Av. Gal. Milton Tavares de Souza s/n, Gragoat\'a, 24210-346, Niter\'oi, RJ, Brazil.}

\date{\today }

\begin{abstract}
We introduce a connection between entanglement induced by interaction and geometric phases acquired by a composite  
quantum spin system. We begin by analyzing the evaluation of cyclic (Aharonov-Anandan) and non-cyclic (Mukunda-Simon) 
geometric phases for general spin chains evolving in the presence of time-independent magnetic fields. Then, by 
considering Heisenberg chains, we show that the interaction geometric phase, namely, the total 
geometric phase with subtraction of free spin contributions, is directly related to  the global (Meyer-Wallach) entanglement 
exhibited by an initially separable state during its evolution in Hilbert space. This is analytically shown for $N=2$ spins 
and numerically illustrated for larger chains. This relationship promotes the interaction geometric phase to an indicator 
of global entanglement in the system, which may constitute a useful tool for quantum tasks based on entanglement as a resource to their performance.
\end{abstract}


\maketitle

\section{Introduction}

Geometric phases have been proposed as a typical mechanism for a quantum system to keep the memory 
of its evolution in Hilbert space. Such phases were introduced in quantum mechanics by Berry 
in 1984~\cite{Berry:84}. Soon after Berry's work, geometric phases became objects of intense theoretical 
and experimental research~\cite{GP-Book:89}, whence it was noticed that they are related to a number 
of important phenomena in physics~\cite{GP-Book:03}, such as Aharonov-Bohm~\cite{Aharonov:59} 
and quantum Hall~\cite{Klitzing:80} effects. In recent years, this interest has also increased due to 
their applicability in quantum-information processing~\cite{Zanardi:99,Jones:00}. 

In his seminal work, Berry considered an arbitrary system governed by a time-dependent Hamiltonian $H[\vec{R}(t)]$, where  
$\vec{R}(t)$ denotes a set of parameters (e.g., a three-dimensional magnetic field). It was then shown that, if the system 
is prepared in a non-degenerate eigenstate of the initial Hamiltonian $H[\vec{R}(0)]$ and is kept on adiabatic  
cyclic evolution in parameter space, then it will return to its initial state accompanied by a phase factor that can be split into
a dynamic contribution, which depends both on energy and evolution time, and a 
geometric contribution, which exclusively depends on the path traversed by the system in parameter space. 
Berry phases were generalized in several contexts. Wilczek and Zee 
considered systems under cyclic and adiabatic evolution dictated by Hamiltonians with degenerate spectra, 
which results in non-Abelian (non-commutative) geometric phases~\cite{Wilczek:84}. Aharonov and Anandan treated 
Abelian phases under cyclic but {\it non-adiabatic} evolution~\cite{Aharonov:87}, which were then extended 
by Anandan for the non-Abelian case~\cite{Anandan:88}. Moreover, formulations of {\it noncyclic} geometric phases were proposed 
by Samuel and Bhandari~\cite{Samuel:88} and by Mukunda and Simon~\cite{MS}. Extensions for mixed states 
under unitary~\cite{Uhlmann:86,Sjoqvist:00} and 
non-unitary~\cite{Garrison:88,Romero:02,Carollo:03,Ericsson:03,Whitney:03,Marzlin:04,Sarandy:0607} 
evolutions have also been considered due to their importance in decohering systems.

Recently, the interface between geometric phases and quantum-information theory has been attracting renewed attention, with focus on 
investigating of the effect of entanglement on the behavior of geometric phases in composite 
systems~\cite{Tong:03,Yi:04,Bertlmann:04,Yi:05,Cui:06,Basu:06,Milman:06,Souza:07,Williamson:07,Niu:10,Sponar:10,Oxman:10}. 
In particular, consider a composite quantum state evolving in time by local unitary transformations. 
Then, the available entanglement is naturally 
kept fixed for the state. For such evolution, the total dynamic phase for the composite state can be split as the 
sum of the dynamic phases for the subsystems. In contrast, this does not hold for the geometric phase unless the 
state is separable~\cite{Williamson:07}. In other words, entangled states acquire a correction term in the 
geometric phase for the composite system with respect to the sum of geometric phases for each subsystem. 
This correction  arises from the (classical and quantum) correlations among the parts of the entangled state.

In this context, the aim of this work is to discuss the relationship between geometric phases and entanglement in a 
composite quantum system in the presence of {\it interaction} among the parts of the system. In this case, we have 
that the evolution will be given by nonlocal unitary transformations (nonfactorizable in the tensor product 
of individual transformations). In particular, if a system is prepared in a separable 
initial state, the interaction may induce the appearance of entanglement during the state evolution. We will be 
interested in the role played by entanglement on the geometric phase acquired during the evolution in comparison 
with the geometric phase acquired by the system in the non-interacting limit. As we will show, for the case of 
Heisenberg spin chains, the contribution of the exchange interaction to the geometric phase allows one to identify product states 
as well as to quantify the global (Meyer-Wallach) entanglement~\cite{meyerwallach} induced during the evolution. 

The work is organized in the following way. In Section II, we present a general approach to determine Aharonov-Anandan (AA)  
and Mukunda-Simon (MS) geometric phases for quantum spin chains in time-independent magnetic fields. In Section III, 
we focus on Heisenberg interactions, obtaining a relationship between geometric phases and global entanglement. 
In Section IV, we summarize our conclusions and discuss the perspectives of future investigations.

\section{Geometric phases for spin systems in time-independent magnetic fields}

\label{generaltreatment}

Let us begin by presenting a systematic procedure for obtaining cyclic and noncyclic geometric phases 
in a quantum spin chain of arbitrary length $N$ immersed in a time-independent magnetic field $\vec{B} = B\hat{z}$. 
We consider that the system is described by a Hamiltonian given by 
\begin{equation}
 H = J H_I + B\displaystyle\sum_{i=1}^{N} \sigma_{i}^{z},
\label{hgeral}
\end{equation}
where $H_I$ is the interaction Hamiltonian among the spins of the lattice, $J$ provides the energy scale for the interacting spins, and 
$B$ denotes the magnetic field. It will be assumed that
\begin{equation}
\left[ H_{I}, S^{z} \right] = 0 ,
\label{u1inv}
\end{equation}
where $S^{z}$ is the operator describing the total spin in the $z$ direction, namely, 
\begin{equation}
 S^{z} = \displaystyle\sum_{i=1}^{N} \sigma_{i}^{z}.   
\end{equation}
Therefore, $H_{I}$ and $S^{z}$ display a simultaneous basis of eigenstates $\{|\varphi_{n}\rangle\}$, i.e.
\begin{eqnarray}
 H_{I}|\varphi_{n}\rangle = E_{n}|\varphi_{n}\rangle, \label{ehi}\\
 S^{z}|\varphi_{n}\rangle = M_{n}|\varphi_{n}\rangle, \label{estot}
\end{eqnarray}
where $E_{n}$ is the energy associated with the eigenstate $|\varphi_{n}\rangle$ and $M_{n}$ its respective magnetization. 
Expanding $|\psi (t)\rangle$ in the common basis of eigenstates of $H_{I}$ and $S^{z}$, we obtain
\begin{equation}
 |\psi(t)\rangle = \displaystyle\sum_{n} c_{n}(t)|\varphi_{n}\rangle .
\label{psiexp}
\end{equation}
Using Eq.~(\ref{psiexp}) into the time-dependent Schr\"odinger equation
\begin{equation}
 H|\psi(t)\rangle = i\hbar|\dot{\psi}(t)\rangle,
\end{equation}
with the dot symbol over $|\psi(t)\rangle$ denoting the derivative with respect to $t$, we get
\begin{equation}
 \displaystyle\sum_{n} c_{n}(t)(JE_{n}+BM_{n})|\varphi_{n}\rangle = i\hbar\displaystyle\sum_{n}\dot{c}_{n}(t)|\varphi_{n}\rangle.
\label{diffEqs}
\end{equation}
Projection of Eq.~(\ref{diffEqs}) on $\langle \varphi_m|$ yields the solution
\begin{equation}
 c_{n}(t) = c_{n}(0) e^{-\frac{i}{\hbar}\left(JE_{n}+BM_{n}\right)t}.
\end{equation}
Hence, from Eq.~(\ref{psiexp}), the state vector $|\psi(t)\rangle$ is given by
\begin{equation} \label{psi}
 |\psi(t)\rangle = \displaystyle\sum_{n} e^{-\frac{i}{\hbar}\left(JE_{n}+BM_{n}\right)t} c_{n}(0)|\varphi_{n}\rangle.
\end{equation}
We rewrite Eq.~(\ref{psi}) in the form
\begin{equation} \label{psi2}
 |\psi(t)\rangle = \displaystyle\sum_{n} e^{i\alpha_{n}(t)} c_{n}(0) |\varphi_{n}\rangle, 
\end{equation}
with
\begin{equation}
 \alpha_{n}(t) = -\frac{1}{\hbar}(JE_{n} + BM_{n})t . 
\label{alphaconst}
\end{equation}
Then, in order to compute the MS geometric phase for the chain at an arbitrary time $t$, we evaluate~\cite{MS}
\begin{equation} \label{fgms2}
 \gamma_{MS} = {\textrm{Arg}}\left[ \langle\psi(0)|\psi(t)\rangle \right] + \frac{1}{\hbar}\int_{0}^{t} dt^\prime\langle\psi(t^\prime)|H|\psi(t^\prime)\rangle.
\end{equation}
For the AA geometric phase~\cite{Aharonov:87}, a further step is still necessary. We have to determine cyclic evolutions for $|\psi(t)\rangle$ 
in projective Hilbert space, whose elements are equivalence classes of state vectors that are equal up to complex phases.  
In this direction, we choose an eigenstate $|\varphi_{n}\rangle$ that is present in the expansion given by Eq.~(\ref{psi2}) and write
\begin{equation}
 |\psi(t) \rangle = e^{i\alpha_{n}} \left[ c_{n}(0)|\varphi_{n}\rangle + \displaystyle\sum_{m \neq n} 
                    e^{i\left[ \alpha_{m}-\alpha_{n} \right]} c_{m}(0) |\varphi_{m}\rangle \right] \, .
\end{equation}
Therefore, a cyclic evolution for $|\psi(t)\rangle$ in projective Hilbert space occurs for a time $\tau$ such that
\begin{equation}
 e^{i\left[ \alpha_{m}(\tau)-\alpha_{n}(\tau)\right]} = e^{i 2\pi p_{nm}} , 
\label{tcic1}
\end{equation}
with $p_{nm}$ $\in$ $\mathbb{Z}$. 
Using Eq.~(\ref{alphaconst}) in Eq.~(\ref{tcic1}), we obtain the period $\tau$ for the cyclic evolution, namely,
\begin{equation} \label{tc}
\tau_{nm} = \frac{2\pi \hbar p_{nm}}{J(E_{n}-E_{m})+B(M_{n}-M_{m})}.
\end{equation}
Naturally, Eq.~(\ref{tc}) must {\it not} be applied for indices $n$ and $m$ such that 
$\alpha_{m}(t) = \alpha_{n}(t)$ for arbitrary time $t$ (such eigenstates are already in phase for any $t$). 
Moreover, observe that $\tau_{nm}$ must be the same for any $n$ and $m$. This is obtained by suitably adjusting the integer numbers $p_{nm}$. 
We will illustrate this procedure in the next section. Then, let us write $\tau \equiv \tau_{nm}$ ($\forall \, n,m$), 
which means that the total phase will be
\begin{equation} \label{ftaa}
 \phi = \alpha_{n}(\tau) = -\frac{JE_{n} + BM_{n}}{\hbar} \tau \, .
\end{equation}
Note that $\phi$ is defined up to $\pm 2\pi q$ ($q \in \mathbb{Z}$). For the dynamic phase after period $\tau$ we get
\begin{eqnarray} \label{fasedaa}
 \phi_{d} &=& -\frac{1}{\hbar} \int_{0}^{\tau} \langle \psi(t)|H|\psi(t)\rangle dt   \nonumber \\
&=& -\frac{1}{\hbar} \displaystyle\sum_{n} |c_{n}(0)|^{2} \left( JE_{n}+BM_{n} \right) \tau \, .
\end{eqnarray}
Hence, the AA geometric phase, which is defined by $\beta = \phi - \phi_{d}$~\cite{Aharonov:87}, can be written as
\begin{equation}
 \beta = \frac{\tau}{\hbar}\left[- \left(JE_{n} + BM_{n}\right) + \displaystyle\sum_{m} |c_{m}(0)|^{2} \left( JE_{m}+BM_{m} \right) \right].
\label{gphasegen}
\end{equation}
Equation~(\ref{gphasegen}) holds for any Hamiltonian $H_I$ obeying Eq.~(\ref{u1inv}) governing chains with arbitrary length $N$. As a general 
application of this result, note that if $|\psi(0)\rangle$ is an eigenstate of $H_I$, i.e. $|\psi(0)\rangle = |\varphi_n\rangle$, 
then the evolution is cyclic for any time $t$ [since $|\psi(0)\rangle$ will be a stationary state of $H$]. In this case, 
the geometric phase acquired by the state vector $|\psi(\tau)\rangle$ will be vanishing (modulo $2\pi$), since 
$\phi=\phi_d$ as given by Eqs.~(\ref{ftaa}) and (\ref{fasedaa}). Therefore, nontrivial geometric phases appear only for initial states that are superpositions containing more than 
one eigenstate of $H_I$. 

\section{Geometric phases and global entanglement for Heisenberg chains}

In this Section, we will focus on the discussion of geometric phases in quantum spin$-\frac{1}{2}$ chains governed by Heisenberg interactions, 
whose $H_I$ reads
\begin{equation}
H_{I}= \sum_{i=1}^{N} \vec{\sigma}_i \cdot \vec{\sigma}_{i+1} = \sum_{i=1}^{N} \left( \sigma^x_i \sigma^x_{i+1} + 
\sigma^y_i \sigma^y_{i+1} + \sigma^z_i \sigma^z_{i+1} \right),
\label{HHeis}
\end{equation}
where periodic boundary conditions are assumed, i.e. $\sigma^\alpha_{N+1}=\sigma^\alpha_{1}$ ($\alpha=x,y,z$). We will first 
discuss AA geometric phases in detail for the two-spin and larger Heisenberg chains, leaving the analysis of MS geometric phases for the 
last section.  

\subsection{Two-spin Heisenberg chain}

Let us start by considering the simple case of a two-spin chain, whose interaction Hamiltonian is simply given by 
$H_I=\vec{\sigma}_1 \cdot \vec{\sigma}_{2}$. At time $t=0$, we 
assume  that the composite system is prepared in an arbitrary separable pure state  $|\psi(0)\rangle$ given by
\begin{equation}
 |\psi(0)\rangle = \bigotimes_{i=1}^2 \left( \cos\frac{\theta_{i}}{2} |+\rangle + e^{i\phi_{i}} \sin \frac{\theta_{i}}{2} |-\rangle \right), 
\label{eiheis}
\end{equation}
where the basis vectors $\{|+\rangle, |-\rangle\}$ are the eigenstates of $\sigma_z$ (computational basis), with $\theta_1$, $\theta_2 \in [0,\pi]$ 
and $\phi_1$, $\phi_2 \in [0,2\pi)$. Evaluating the tensor product in Eq.~(\ref{eiheis}), we obtain
\begin{equation}
 |\psi(0)\rangle = a_{1}(0)|++\rangle + a_{2}(0)|+-\rangle + a_{3}(0)|-+\rangle + a_{4}(0)|--\rangle , 
\label{eiasheis2}
\end{equation}
with
\begin{eqnarray}
 a_{1}(0) &=& \cos \frac{\theta_{1}}{2} \cos \frac{\theta_{2}}{2}, \nonumber \\
 a_{2}(0) &=& e^{i\phi_{2}} \cos \frac{\theta_{1}}{2} \sin \frac{\theta_{2}}{2}, \nonumber \\
 a_{3}(0) &=& e^{i\phi_{1}} \cos \frac{\theta_{2}}{2} \sin \frac{\theta_{1}}{2}, \nonumber \\
 a_{4}(0) &=& e^{i(\phi_{1}+\phi_{2})} \sin \frac{\theta_{1}}{2} \sin \frac{\theta_{2}}{2}.
\end{eqnarray}
Diagonalizing $H_{I}$, we get the eigenvectors
\begin{eqnarray}
 |\varphi_{1}\rangle &=& |++\rangle \, , \hspace{0.5cm} E_{1} = +1 \, , \hspace{0.5cm} M_{1} = +2\, , \nonumber \\
 |\varphi_{2}\rangle &=& |--\rangle \, , \hspace{0.5cm} E_{2} = +1 \, , \hspace{0.5cm} M_{2} = -2 \, ,\nonumber \\
 |\varphi_{3}\rangle &=& \frac{1}{\sqrt{2}} \left(|+-\rangle + |-+\rangle \right) \, , \hspace{0.18cm} E_{3} = +1 \, , \hspace{0.5cm} M_{3} = 0 \, , 
\nonumber \\
 |\varphi_{4}\rangle &=& \frac{1}{\sqrt{2}} \left(|+-\rangle - |-+\rangle \right) \, , \hspace{0.18cm} E_{4} = -3 \, , \hspace{0.5cm} M_{4} = 0 \, , 
\nonumber \\
\end{eqnarray}
with $E_i$ and $M_i$ ($i=1,\cdots,4$) denoting their energies and magnetizations, which are defined by Eqs.~(\ref{ehi}) and~(\ref{estot}), respectively. Rewriting $|\psi(0)\rangle$ in terms of the eigenstates of $H_{I}$, we have
\begin{equation}
 |\psi(0)\rangle = c_{1}(0)|\varphi_{1}\rangle + c_{2}(0)|\varphi_{2}\rangle + c_3 (0) |\varphi_{3}\rangle + c_4 (0) |\varphi_{4}\rangle, 
\end{equation}
with
\begin{eqnarray}
 c_{1}(0) &=& a_{1}(0) , \nonumber \\
 c_{2}(0) &=& a_{4}(0) , \nonumber \\
 c_{3}(0) &=& \frac{a_{2}(0) + a_{3}(0)}{\sqrt{2}} , \nonumber \\
 c_{4}(0) &=& \frac{a_{2}(0) - a_{3}(0)}{\sqrt{2}} . 
\end{eqnarray}
Observe that, from the energies $E_n$, magnetizations $M_n$, and amplitudes $c_n(0)$, we obtain the MS and AA geometric phases 
directly from Eqs.~(\ref{fgms2}) and (\ref{gphasegen}), respectively. 
Let us investigate AA geometric phases and take $0 < (\theta_1$, $\theta_2) < \pi$ (i.e., $\theta_1$ and $\theta_2$ are not on  
the boundary). Then the eigenstate $|\varphi_1\rangle$ of 
$H_I$ is present in the expansion of the initial state $|\psi(0)\rangle$. Therefore, we can write out $|\psi(t)\rangle$ as
\begin{equation}
 |\psi(t)\rangle = e^{i\alpha_{1}(t)} \left[ c_{1}(0)|\varphi_{1}\rangle + \displaystyle\sum_{m=2}^{4} c_{m}(0) 
                  e^{i(\alpha_{m}-\alpha_{n})}|\varphi_{m}\rangle \right].
\label{psitheis2}
\end{equation}
Using now Eq.~(\ref{tc}), we determine the expressions for the cyclic time
\begin{eqnarray}
\tau_{12} &=& \frac{\pi \hbar}{2B} p_{12}, \\
\tau_{13} &=& \frac{\pi \hbar}{B} p_{13}, \\
\tau_{14} &=& \frac{\pi \hbar}{2J + B} p_{14}.
\end{eqnarray}
As discussed before, all the expressions above for $\tau_{mn}$ must be equivalents, so that $\tau_{12} = \tau_{13} = \tau_{14}$. 
Then we conclude that 
\begin{equation}
p_{12} = 2 p_{13}, \hspace{1cm} p_{14} = \frac{2J + B}{B} p_{13}.
\label{pconditions}
\end{equation}
Assuming that $J$ and $B$ are rational numbers ($J$,~$B \in \mathbb{Q}$), we have that it is always possible to find 
integers $p_{12}$, $p_{13}$, and $p_{14}$ such that Eq.~(\ref{pconditions}) is satisfied. Indeed, if we take 
$\frac{2J + B}{B}=\frac{k}{l}$ (for $l$, $k \in \mathbb{Z}$) then we have that the first period (cyclic time) 
can be obtained by choosing $p_{13}=l$, $p_{14}=k$, and $p_{12}=2 \, l$ (longer cyclic times can be obtained by choosing 
$p_{13}$ as a multiple of $l$ and conveniently adjusting $p_{12}$ and $p_{14}$). Therefore, we can arbitrarily express 
the cyclic time as 
\begin{equation}
\tau = \frac{\pi \hbar}{B} p , 
\label{tcheis2}
\end{equation}
where $p \equiv p_{13}$, with $p$ an integer number that is a function of $J$ and $B$. To compute the total phase $\phi$, 
we use Eq.~(\ref{ftaa}) taking $\alpha_n(t) \equiv \alpha_1(t)$, which yields
\begin{equation}
\phi  =  -\frac{JE_{1} + BM_{1}}{\hbar}\, \tau  =  -2\pi p \left( \frac{J}{2B} + 1 \right) .
\label{ftheis2}
\end{equation}
By using the values above for $c_{n}(0)$, $E_{n}$, $M_{n}$, and $\tau$ into Eq.~(\ref{fasedaa}), the dynamic phase becomes
\begin{eqnarray}
 \phi_d  &=&  - \frac{\pi \,p}{B} \left[ B \left( \cos\theta_1 + \cos\theta_2 \right) + J \cos\theta_1 \cos\theta_2 \right. \nonumber \\
&&\left. + J \cos(\phi_1-\phi_2) \sin\theta_1 \sin\theta_2 \right] .
\label{fdheis2}
\end{eqnarray}
Hence, for the AA geometric phase, we have $\beta = \phi - \phi_d$,  which results in
\begin{equation}
 \beta = \beta_{F} + \beta_{I}, 
\label{betaheis2}
\end{equation}
where
\begin{equation}
 \beta_{F} = -\pi p \left [ \,(1-\cos\theta_{1}) + (1-\cos\theta_{2}) \, \right]
\end{equation}
and
\begin{equation}
\beta_{I} = -\frac{J\pi p}{B} \left[ 1 - \cos \theta_{1} \cos \theta_{2} - \cos (\phi_{1}-\phi_{2}) \sin \theta_{1}\sin \theta_{2} \right].
\label{betaintheis2}
\end{equation}
Note that $\beta_{F}$ is related to the geometric phase acquired by the system composed by two {\it free} spins$-\frac{1}{2}$ particles, 
which is basically given by one half of the solid angle described by each particle during its cyclic evolution in the projective space. 
On the other hand, $\beta_{I}$ corresponds to the contribution of the exchange interaction $J$ to the AA geometric phase. Observe that, 
for some states, the interaction geometric phase $\beta_{I}$ will be vanishing. For example, by considering states such that $\theta_1=\theta_2$ 
and $\phi_1=\phi_2$, we will have $\beta_{I}=0$. This means that spins initially prepared in the same direction behave, concerning their 
geometric phases, as if they were free particles.
Observe also that, as indicated just before Eq.~(\ref{psitheis2}), Eqs.~(\ref{betaheis2})-(\ref{betaintheis2}) for the AA geometric phase 
do not automatically apply to the boundary values $0$ and  $\pi$ for $\theta_1$ and $\theta_2$. In these cases, some of the eigenstates 
$|\varphi_n\rangle$ of $H_I$ will not be present into Eq.~(\ref{eiheis}) for the initial state, which can modify both the cyclic time 
given by Eq.~(\ref{tcheis2}) and the total phase given by Eq.~(\ref{ftheis2}). 
From a general point of view, the boundaries are determined by a gauge choice. In fact, the representation adopted for each spin 
in Eq.~(\ref{eiheis}) defines the angular variables so that the poles of the Bloch sphere are in the $\mathit{z}$ axis. Such poles can be 
changed by a gauge transformation, which implies that the boundary is an artifact that is not essential in our treatment (similar 
to the Dirac string singularity of magnetic monopoles). 
For boundary values, a case by case analysis is demanded, which, 
however, will not spoil the relationship between $\beta_I$ and entanglement to be introduced next.
 
Indeed, we can show that the interaction geometric phase for the Heisenberg chain, given either by Eq.~(\ref{betaintheis2}) for two spins 
or numerically obtained for chains of length $N$, can be related to multipartite entanglement as measured by the global entanglement 
introduced by Meyer and Wallach in Ref.~\cite{meyerwallach}. Given a state vector $|\psi\rangle$ describing a pure composite quantum 
system containing $N$ qubits, the global entanglement $Q$ can be determined by~\cite{brennen}   
\begin{equation}
 Q\left( |\psi\rangle \right) = 2\left[1 - \frac{1}{N} \displaystyle\sum_{k=1}^{N} Tr\left(\rho_{k}^{2}\right)\right], 
\label{qmw}
\end{equation}
where $\rho_{k}$ is the reduced density matrix for qubit $k$. The global entanglement is normalized such that 
$0 \leq Q \leq 1$, with $Q=0$ if, and only if, $|\psi\rangle$ is a product state, 
and $Q=1$ for entangled states such that $Tr\left(\rho_{k}^{2}\right)=1/2, \,\,\forall k$, i.e, for vectors $|\psi \rangle$ 
such that the reduced state of each qubit is a maximally mixed state. In order to compute $Q\left( |\psi\rangle \right)$ for 
the Heisenberg model, we write the single-site density operator at site $k$ as
\begin{equation}
\rho_{k} = \frac{1}{2} \left( \mathds{1}_k + \sum_{\alpha=x,y,z} v_k^\alpha \sigma_k^\alpha \right), 
\label{rhokheis}
\end{equation}
where $v_k^\alpha$ denotes the coherence vector for site $k$, which is given by $v_k^\alpha = \langle \sigma_k^{\alpha} \rangle$, 
with the expectation value $\langle \sigma_k^{\alpha} \rangle$ taken with respect to the state vector $|\psi(t)\rangle$ provided 
by Eq.~(\ref{psi2}). Analytical computation for an arbitrary time $t$ evolved from a separable state as in Eq.~(\ref{eiheis}) yields
\begin{eqnarray}
 Q &=& \frac{1}{4} \sin^2 \left(\frac{4Jt}{\hbar}\right) \left[ 1 - \cos \theta_{1}\cos \theta_{2} \right. \nonumber \\
&&\left. -\cos \left(\phi_{1}-\phi_{2}\right) \sin\theta_{1} \sin\theta_{2}\right]^{2}.
\label{EmaranhamentoGlobal}
\end{eqnarray}
However, for a cyclic evolution, the system will return (up to a phase) to its original state. 
Then, starting from a separable state, entanglement vanishes both at the beginning and at the end of evolution. Therefore, 
in order to compare global entanglement with the AA geometric phase, it is useful to consider the average global entanglement after 
period $\tau$. From Eqs.~(\ref{betaintheis2}) and (\ref{EmaranhamentoGlobal}), we obtain a remarkable relationship between 
the average global entanglement $Q_{med}$ and the square of the interaction geometric phase $\beta_I$, which reads    
\begin{equation}
 Q_{med}\left(|\psi\rangle\right) = \left( \frac{B}{2\pi pJ}\right)^{2} g(\tau) \, \, \beta_{I}^{2},
\end{equation}
with
\begin{equation}
g\left(\tau\right) = \frac{1}{\tau} \int_{0}^{\tau} \sin^{2}\left(\frac{4Jt}{\hbar}\right) dt = 
\left(\frac{1}{2}-\frac{\sin(\frac{8 J \tau}{\hbar})}{16 J \tau/\hbar}\right).
\end{equation}
Therefore, the interaction contribution $\beta_{I}$ for the AA geometric phase is able to reveal the global entanglement induced by the 
Heisenberg exchange coupling $J$ for an initially separable state. 

\subsection{Larger Heisenberg chains}

\begin{figure}
 \centering
\includegraphics[scale=0.26]{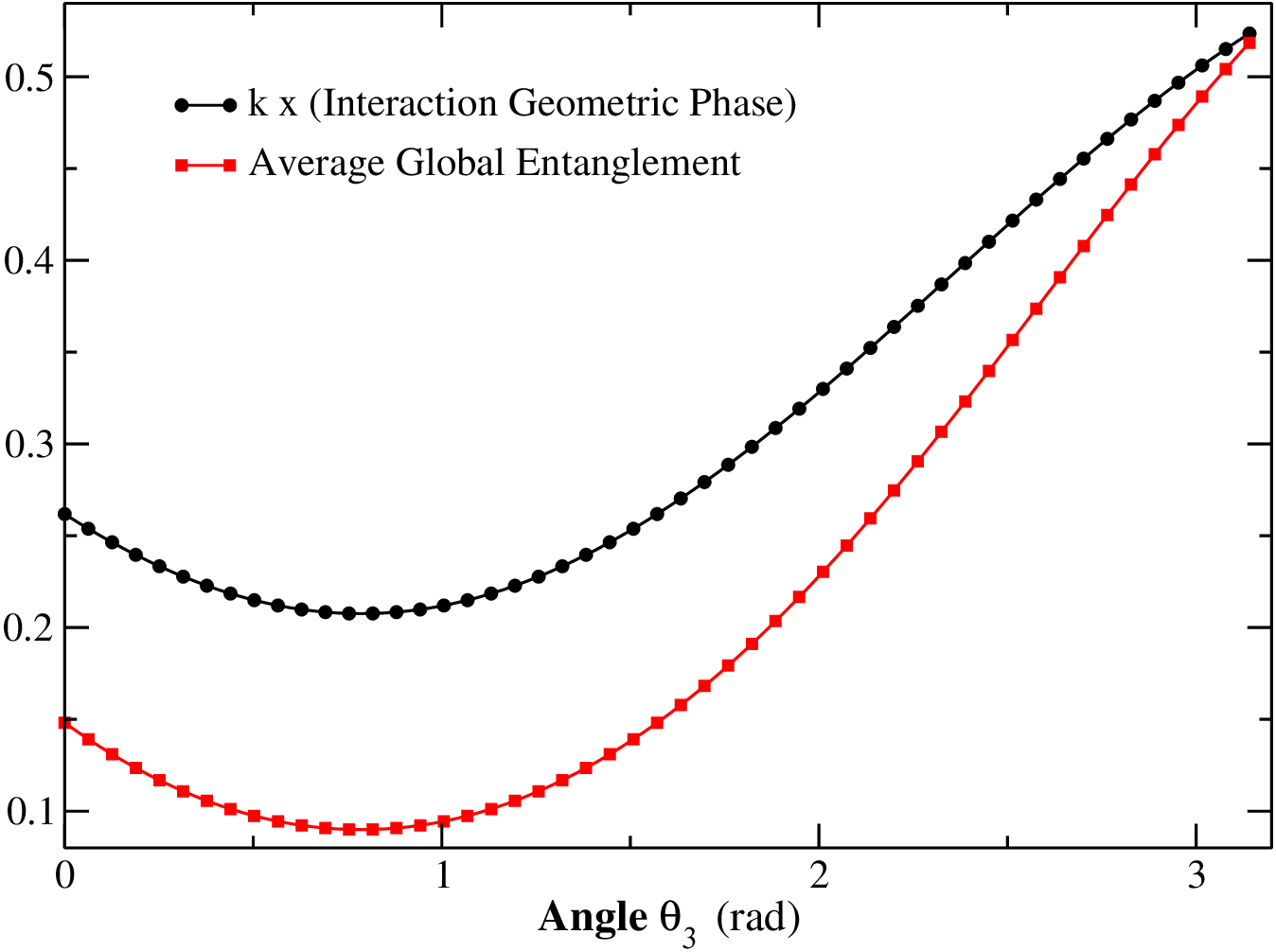}
\caption{{\scriptsize{(Color online) Interaction AA geometric phase $\beta_{I}$, multiplied by a constant $k$, and average global 
entanglement as a function of the angle $\theta_{3}$ for the three-spin Heisenberg chain in a constant magnetic field with 
initial state given by Eq.~(\ref{estado001T}). 
We adopt $\hbar = 1$ and $B/J=3$.}}}
\label{fig33}
\end{figure}

\begin{figure}
 \centering
\includegraphics[scale=0.26]{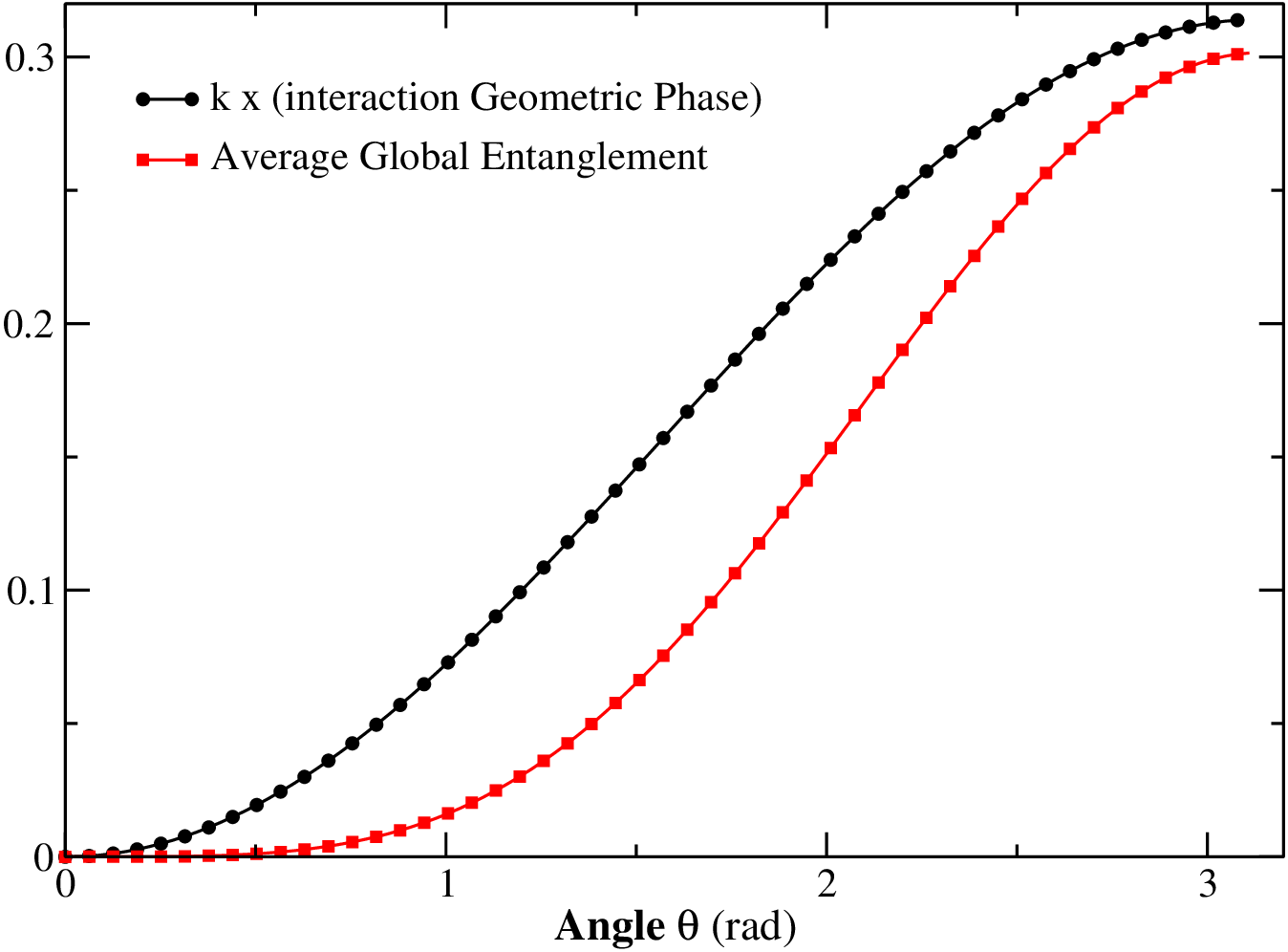}
\caption{{\scriptsize{(Color online) 
Interaction AA geometric phase $\beta_{I}$, multiplied by a constant $k$, and average global entanglement as a function of the 
angle $\theta$ for the Heisenberg chain with $N=12$ spins in a constant magnetic field with initial state given by Eq.~(\ref{12spins}). 
We adopt $\hbar = 1$ and $B/J=3$.}}}
\label{graf12spins}
\end{figure}

The connection between interaction geometric phases and global entanglement can also be illustrated for larger 
chains, where the necessity of a multipartite measure of entanglement turns out to be explicit. 
Indeed, consider a Heisenberg chain with $N$ sites in a general initial state $|\psi(0)\rangle$ given by
\begin{equation}
 |\psi(0)\rangle = \bigotimes_{i=1}^{N} \left( \cos \frac{\theta_i}{2}|+\rangle + e^{i\phi_i} \sin\frac{\theta_i}{2}|-\rangle\right).
 \label{Nspins}
\end{equation}
As a first illustration, let us take a three-spin chain, whose initial state $|\psi\left(0\right)\rangle$ in Eq.~(\ref{Nspins}) is such that 
$\theta_{1}=0$, $\theta_{2} = \pi /2$, and $\phi_{1} = \phi_{2} = \phi_{3} = 0$.  Then,
\begin{equation}
 |\psi\left(0\right)\rangle = |+\rangle \otimes \left( \frac{|+\rangle  + |-\rangle}{\sqrt{2}}\right)\otimes
                       \left( \cos\frac{\theta_{3}}{2}|+\rangle + \sin\frac{\theta_{3}}{2}|-\rangle \right).
\label{estado001T}
\end{equation}
The period of evolution is a function of $J$ and $B$. In order to consider a concrete case, we take $B/J=3$. Then, by following 
the steps delineated in Section~\ref{generaltreatment}, it is possible to show that the cyclic time is  
$\tau = \frac{\pi \hbar}{B} p$, with $p \in \mathbb{Z}$. For the interaction geometric phase we analytically obtain
\begin{equation}
 \beta_{I} = \pi +\frac{\pi}{3} \left(\cos \theta_{3} + \sin \theta_{3} \right).
\end{equation}
From Eq.~(\ref{estado001T}), we can obtain the evolved state $|\psi(\tau)\rangle$ given by Eq.~(\ref{psi}), whence it directly follows 
the average global entanglement $Q_{med}$. Then, by plotting $ \beta_{I}$ (multiplied by a constant $k$) and $Q_{med}$ in Fig.~\ref{fig33} 
as a function of the angle $\theta_3$, a monotonic relationship is apparent. 
This result turns out to be found for chains with more sites. As an example, let us consider a Heisenberg chain with $N=12$ sites and initial state 
$|\psi(0)\rangle$ given by 
\begin{equation}
 |\psi(0)\rangle = \bigotimes_{i=1}^{11} |+\rangle_{i} \otimes \left( \cos \frac{\theta}{2}|+\rangle + \sin\frac{\theta}{2}|-\rangle\right).
 \label{12spins}
\end{equation}
By numerical diagonalization of $H_I$, we can determine the AA interaction phase $\beta_{I}$ and the average global entanglement 
$Q_{med}$, whose result is exhibited in Fig.~\ref{graf12spins}. Note that $\beta_{I}$ keeps reflecting the 
entanglement induced by the exchange coupling $J$. 
 
\subsection{MS geometric phases and global entanglement}

We can also establish the connection between $\beta_I$ and global entanglement for a noncyclic time $t$. In this direction, 
let us consider now MS geometric phases and begin by investigating a two-spin Heisenberg chain. Then, 
Eq.~(\ref{EmaranhamentoGlobal}) for global entanglement keeps valid. For the geometric phase, we consider the initial state
\begin{equation}
|\psi(0)\rangle = |+\rangle\otimes \left( \cos\frac{\theta_{2}}{2}|+\rangle + \sin\frac{\theta_{2}}{2}|-\rangle \right) .
\label{initialstate}
\end{equation}
We can obtain the MS interaction geometric phase by using Eq.~(\ref{fgms2}) [with $|\psi(t)\rangle$ given by Eq.~(\ref{psi})] 
and subtracting the free geometric phase of each spin. We adopt from now on units such that $\hbar=1$ and $J=1$. 
For the total phase, we get 
\begin{equation}
 \phi = \textrm{Arg} \left[ e^{-i(1+2B)t}\left(1 + \cos\theta_{2} + e^{2iBt}\left( 1 + e^{4it}\right) \sin^{2}\frac{\theta_{2}}{2}
         \right) \right],
\end{equation}
while the dynamic phase reads
\begin{equation}
 \phi_{d} = -t\left(B + (1+B)\cos \theta_{2}\right).
\end{equation}
The free geometric phase of each spin is
\begin{equation}
 \gamma_{MS}^{(j)} = \textrm{Arg}\left[e^{-iBt}\cos^{2}\frac{\theta_{j}}{2} + e^{iBt}\sin^{2}\frac{\theta_{j}}{2}\right] + Bt\cos \theta_{j},
\end{equation}
where $j=1,2$ labels each of the two spins, with $\theta_1=0$. Hence, the MS interaction geometric phase is 
\begin{equation}
 \gamma_{MS}^{INT}(t) = \phi - \phi_{d} -\displaystyle\sum_{j=1}^{2} \gamma_{MS}^{(j)}.
\end{equation}

\vspace{0.1cm}

\begin{figure}[!ht]
 \centering
 \includegraphics[scale=0.265]{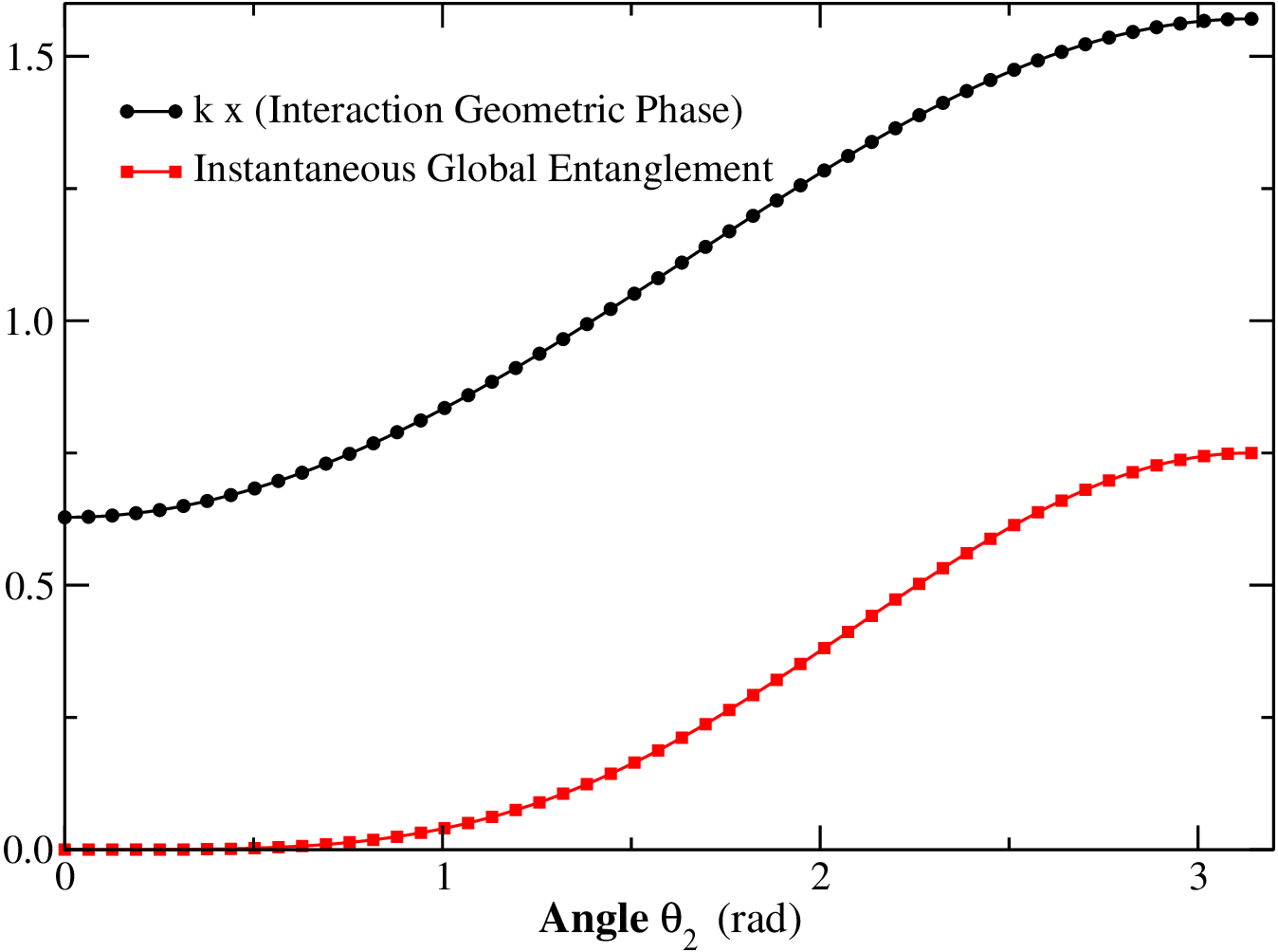}
 \caption{{\scriptsize{(Color online) Interaction geometric phase $\gamma_{MS}^{INT}(t)$, multiplied by a constant $k$, and instantaneous global entanglement as a function of the angle $\theta_{2}$ for the two-spin Heisenberg chain in a constant magnetic field with initial state given by 
Eq.~(\ref{initialstate}). The time is fixed at $t = \pi /3$. We adopt $\hbar=1$ and $B/J = 3$.}}}
\label{f35}
\end{figure}

\vspace{0.1cm}

\begin{figure}[!ht]
 \centering
 \includegraphics[scale=0.265]{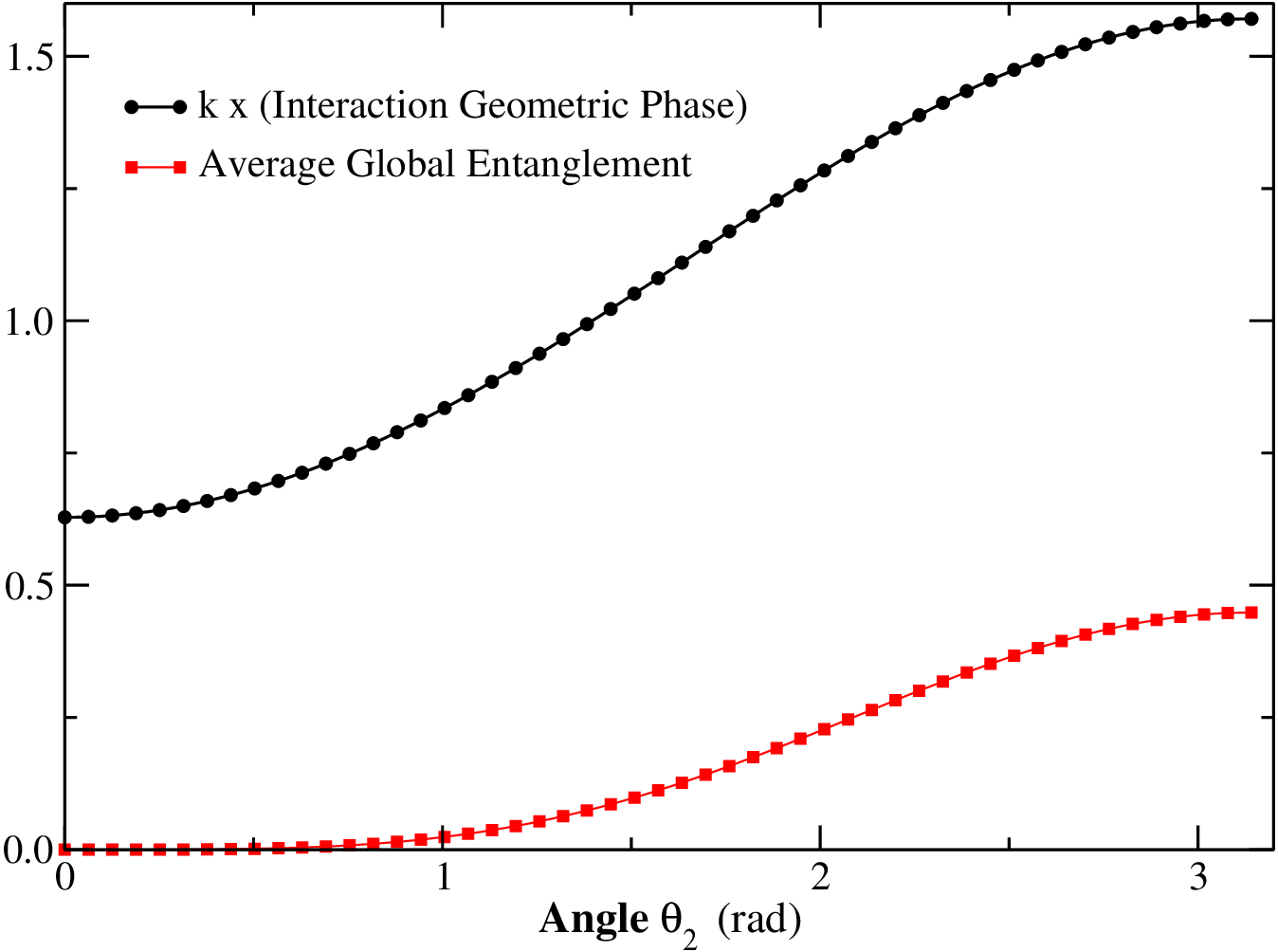}
 \caption{{\scriptsize{(Color online) Interaction geometric phase $\gamma_{MS}^{INT}(t)$, multiplied by a constant $k$, and average global entanglement 
as a function of the angle $\theta_{2}$ for the two-spin Heisenberg chain in a constant magnetic field with initial state given by 
Eq.~(\ref{initialstate}). The time is fixed at $t = \pi /3$. We adopt $\hbar=1$ and $B/J = 3$.}}}
\label{f36}
\end{figure}

Figs.~\ref{f35} and~\ref{f36} illustrate a monotonic relation between the interaction MS phase and both instantaneous and average global 
entanglement, respectively, where time is fixed at $t = \pi /3$. In all those cases, $\gamma_{MS}^{INT}(t)$ reflects the entanglement 
dynamics. Therefore, for the MS geometric phase, the average over entanglement is not a necessary operation (as it was for the AA phase) to 
give evidence of the relation between entanglement and the geometric phase. 

\begin{figure}[ht]
 \begin{minipage}[b]{0.47\linewidth}
  \includegraphics[width=\linewidth]{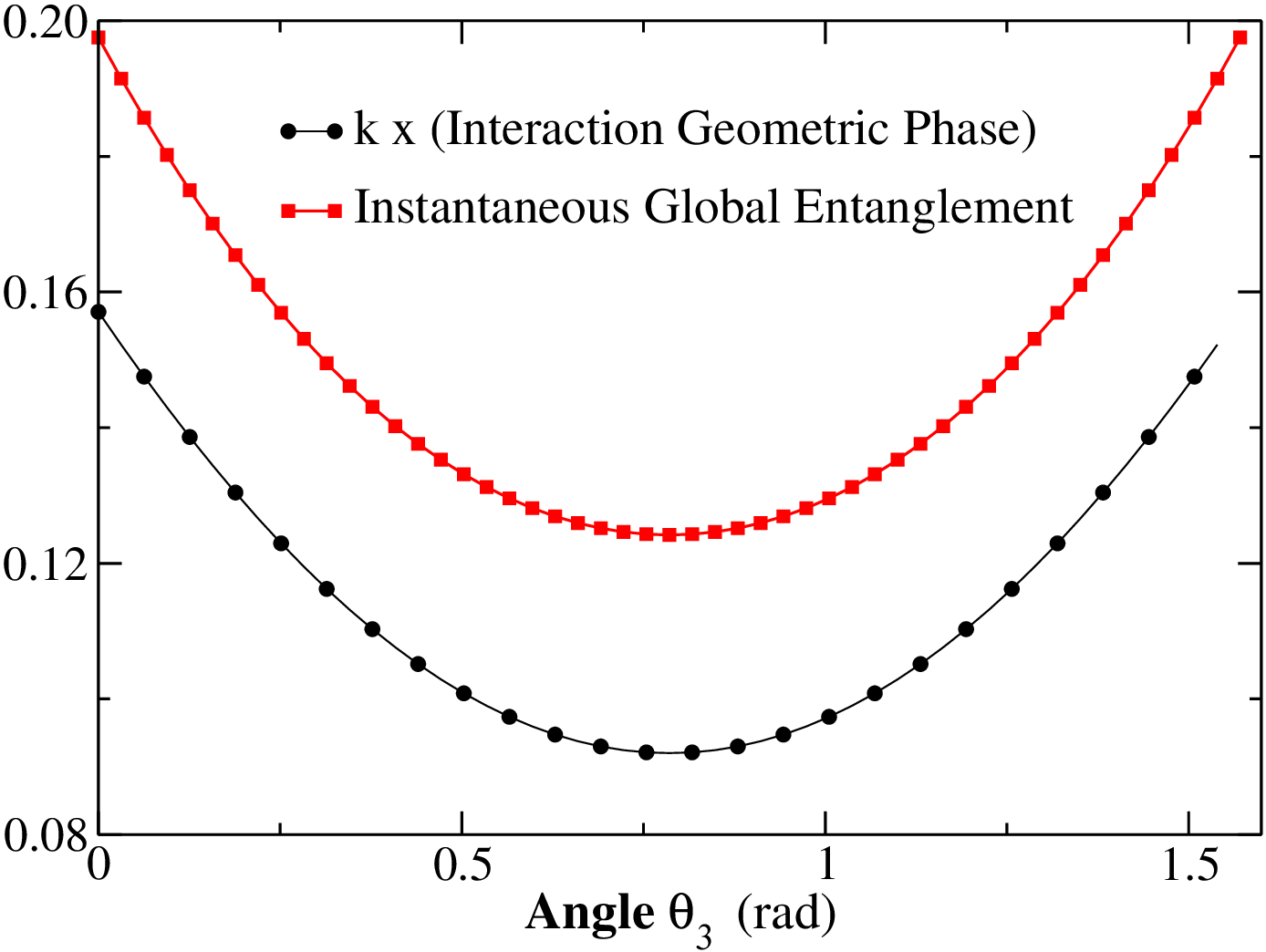}
 \end{minipage} \hfill
\begin{minipage}[b]{0.47\linewidth}
 \includegraphics[width=\linewidth]{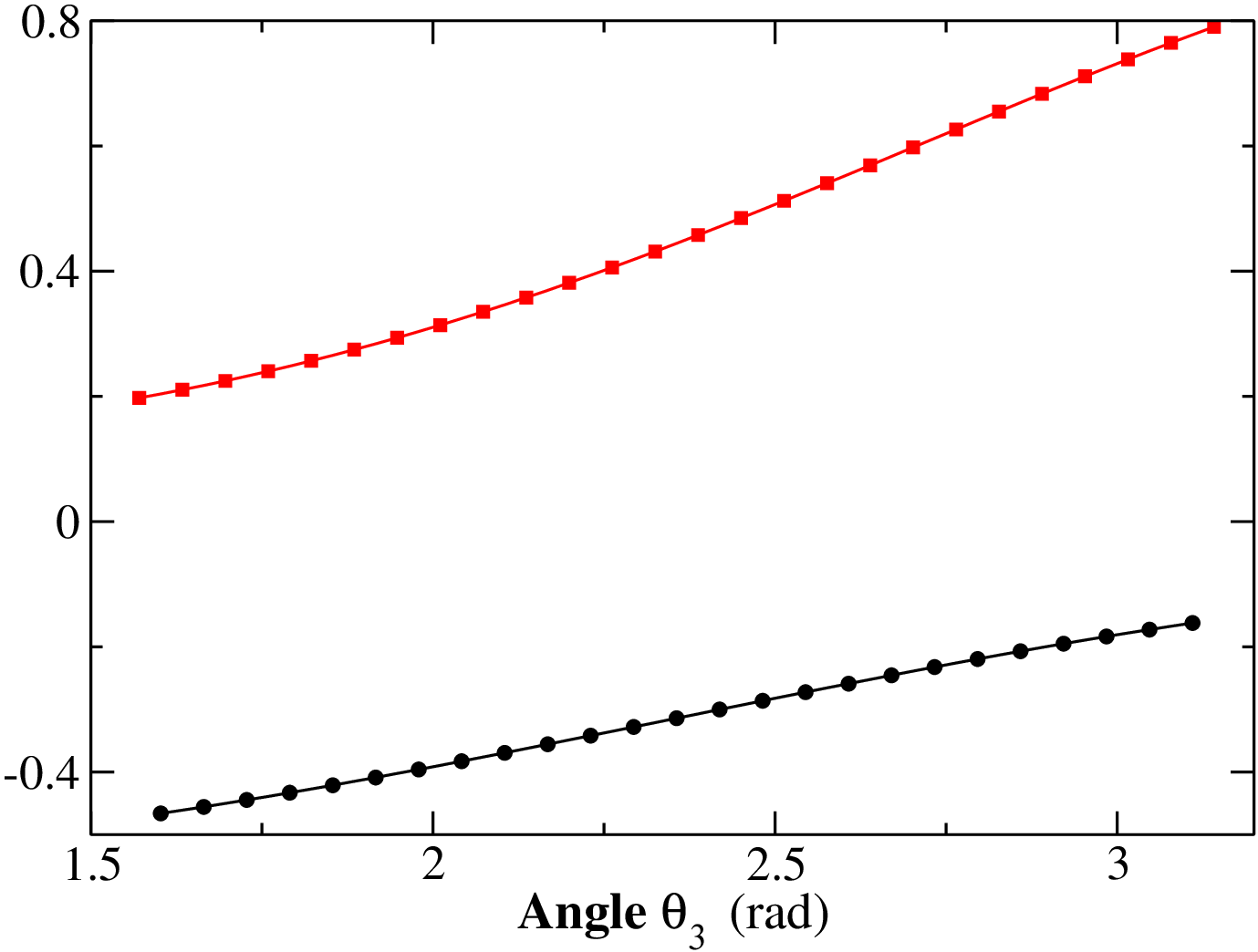}
\end{minipage}
 \caption{{\scriptsize{(Color online) Interaction geometric phase $\gamma_{MS}^{INT}(t)$, multiplied by a constant $k$, and instantaneous global entanglement 
as a function of the angle $\theta_{3}$ for the three-spin Heisenberg chain in a constant magnetic field with initial state given by 
Eq.~(\ref{gpintms}). The time is fixed at $t = \pi /2$. We adopt $\hbar=1$ and $B/J = 3$.}}}
\label{Fig37p1}
\end{figure}

\begin{figure}[ht]
 \begin{minipage}[b]{0.47\linewidth}
  \includegraphics[width=\linewidth]{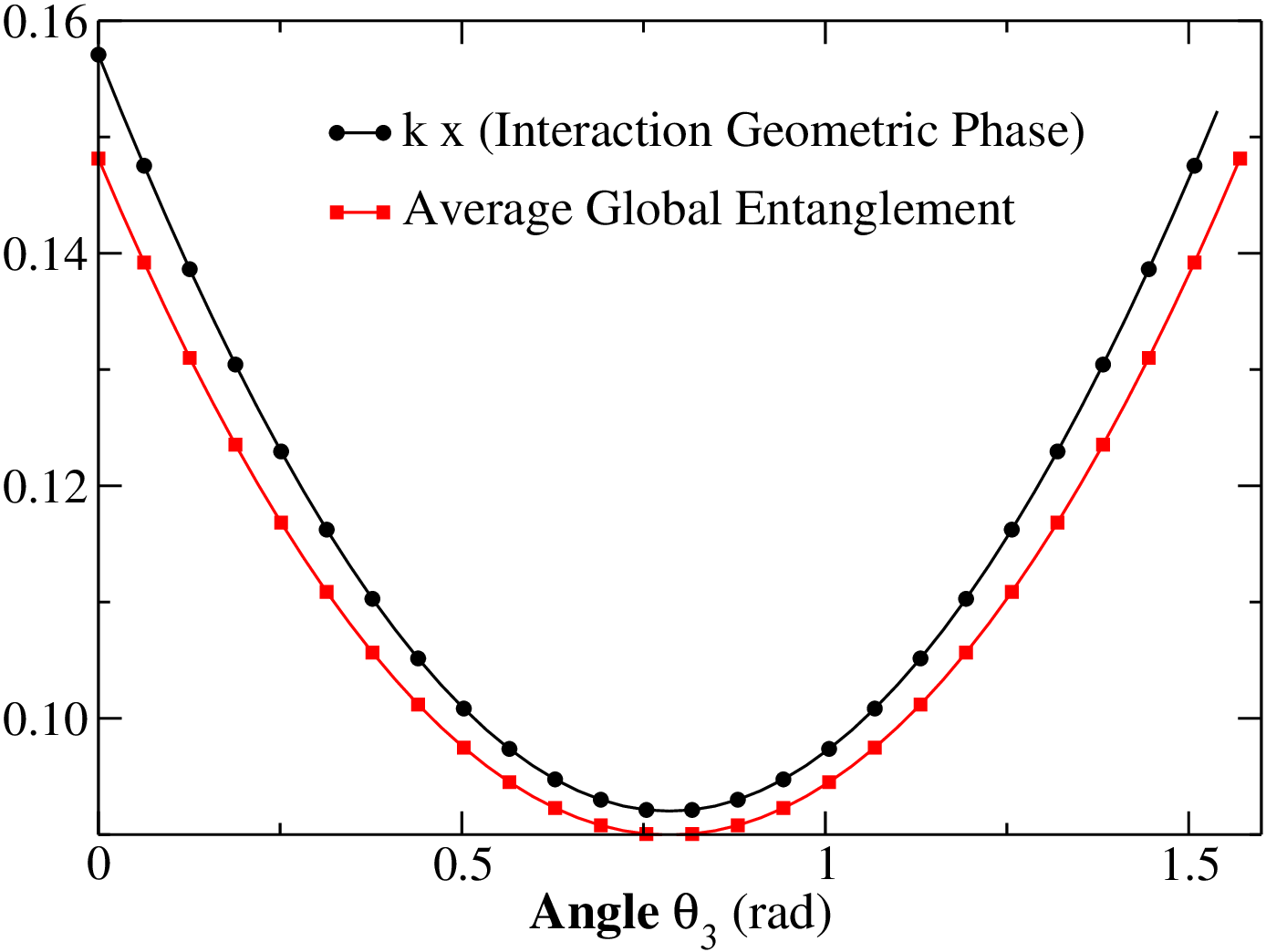}
 \end{minipage} \hfill
\begin{minipage}[b]{0.47\linewidth}
 \includegraphics[width=\linewidth]{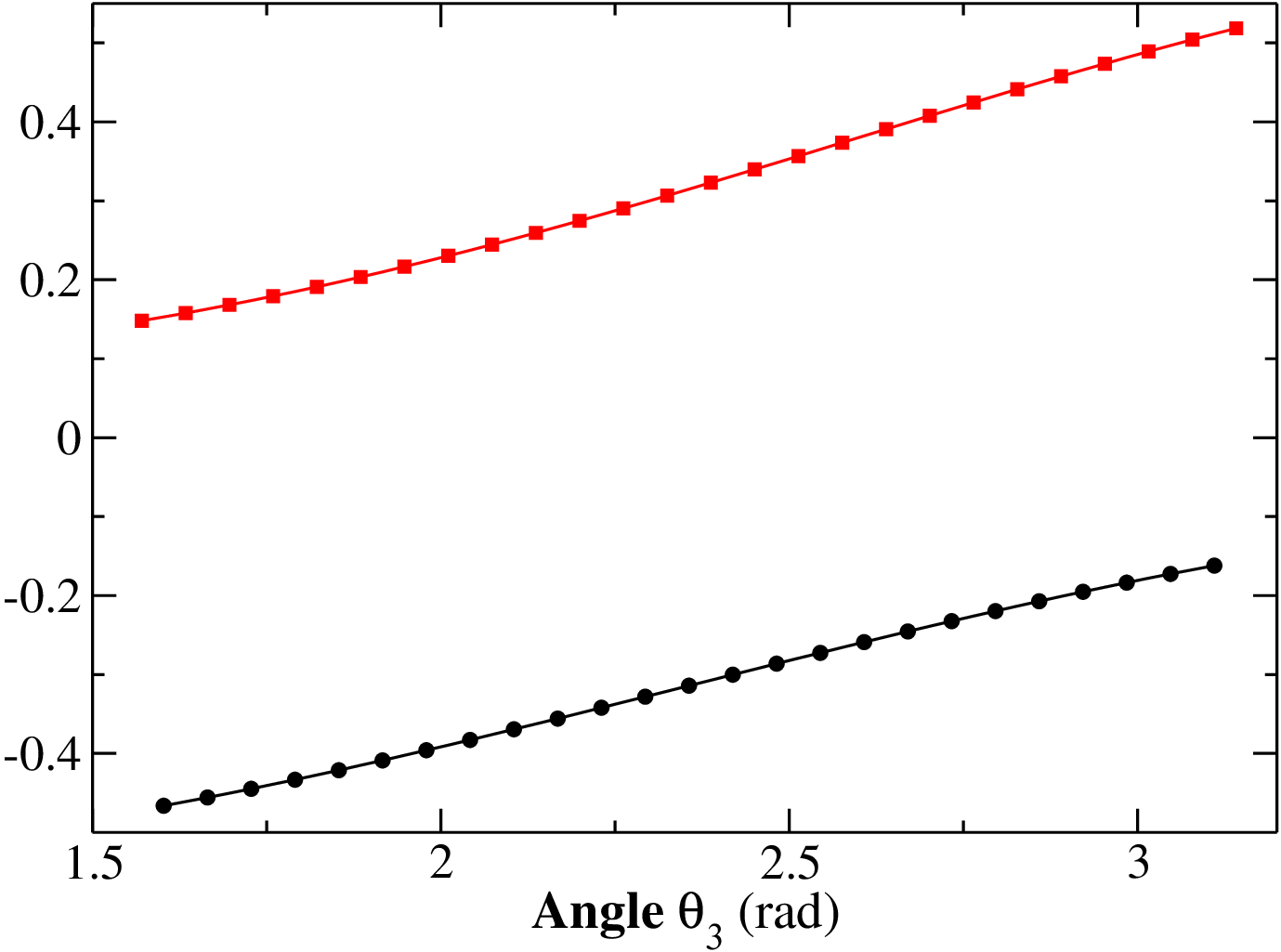}
\end{minipage}
 \caption{{\scriptsize{(Color online) Interaction geometric phase $\gamma_{MS}^{INT}(t)$, multiplied by a constant $k$, and average global entanglement 
as a function of the angle $\theta_{3}$ for the three-spin Heisenberg chain in a constant magnetic field with initial state given by 
Eq.~(\ref{gpintms}). The time is fixed at $t = \pi /2$. We adopt $\hbar=1$ and $B/J = 3$.}}}
  \label{Fig39p1}
\end{figure}

Let us illustrate this result in a multipartite system. In this direction, consider a three-spin Heisenberg chain whose initial state is 
\begin{equation}
 |\psi(0)\rangle = |+\rangle \otimes \left( \frac{|+\rangle + |-\rangle}{\sqrt{2}} \right) \otimes
                   \left( \cos \frac{\theta_{3}}{2} |+\rangle + \sin \frac{\theta_{3}}{2} |-\rangle \right).
 \label{gpintms}
\end{equation}
For this state we fix the time at $t = \frac{\pi}{2}$ and vary the angle $\theta_{3}$. Different from the previous example, 
the MS geometric phase displays now a jump. For $J=1$ and $B=3$, a jump occurs near $\theta_{3} = \pi /2$. 
So we separately plot the interaction MS phase in two parts: first, the angle $\theta_{3}$ varies from $0$ to $\pi /2$, 
and then from $\pi /2$ to $\pi$. This is shown in comparison with the instantaneous global entanglement 
in Fig.~\ref{Fig37p1} and in comparison with the average global entanglement in Fig.~\ref{Fig39p1}. 
From those plots, the monotonicity relation is also apparent. 
\vspace{0.4cm}

\section{Conclusions}

We have introduced a relationship between geometric phases and entanglement in quantum spin systems. 
More specifically, we have analyzed AA and MS phases for Heisenberg chains in time independent magnetic fields, obtaining 
a monotonic relationship between the interaction geometric phase and the global (Meyer-Wallach) entanglement exhibited by the state. 
This relationship promotes the interaction geometric phase to an indicator of the entanglement available in the system, which may 
constitute a useful tool for quantum tasks based on entanglement to their performance. 
Although the detailed reason behind this correspondence is still unresolved in general, a hint to bridge this gap comes from 
the observation that entanglement as well as geometric phases contain information about the correlations of a composite system. 
As an illustration, changes in the correlations reflecting quantum phase transitions~\cite{Sachdev:Book} have been shown to be 
captured by both quantities (see, e.g., Refs.~\cite{Amico:08} and~\cite{Carollo:05}). Further investigation in this direction 
may provide an appealing route for establishing a more definite connection between entanglement and geometric phases. 

Applications of geometric phases to the investigation of correlations among the parts of a quantum system have been pointed out 
in a number of works in recent years. For systems evolving under {\it local transformations}, both separable~\cite{Tong:03,Cui:06,Williamson:07,Niu:10} 
and entangled~\cite{Bertlmann:04,Basu:06,Sponar:10,Oxman:10} states have been shown to display a reflection of their behavior into the geometric phases. 
However, for {\it interacting} systems, where initially separable states may evolve to entangled states and vice versa, the general 
relationship between geometric phases and entanglement is still an open problem. When interaction does {\it not} induce entanglement, 
it has been shown in Ref.~\cite{Niu:10} for the specific case of a two-spin Heisenberg chain that geometric phases are able to indicate 
the separability of a quantum state. In this work, we have considered the more general situation where interaction induces entanglement in the 
Heisenberg chain with $N$ spins, identifying a multipartite measure of entanglement (global entanglement) that is directly related to the 
interaction geometric phase. The relationship between entanglement and geometric phases 
for more general couplings 
and for initially separable mixed states
would be a relevant further contribution to the subject. 
Moreover, it would also be interesting to investigate whether or not {\it initially} nonseparable states may have their 
entanglement somewhat related to the behavior of geometric phases acquired during the quantum evolution. 
We leave these topics to future research.

\section*{Acknowledgements} 

We gratefully acknowledge financial support from
CNPq, FAPERJ, and the Brazilian National Institute for Science and Technology 
of Quantum Information (INCT-IQ).

\end{document}